\documentclass[twocolumn, showpacs, showkeys, preprintnumbers, amsmath,
amssymb, prb, groupedaddress,floatfix]{revtex4}

\usepackage{graphicx}
\usepackage{color}
\usepackage{float}

\begin{document}

\title{Electronic Structure, Localization and
 Spin-State Transition in Cu-substituted FeSe: Fe$_{1-x}$Cu$_x$Se}

\author{Stanislav Chadov}
\email{chadov@uni-mainz.de}
\author{Daniel Sch\"{a}rf}
\author{Gerhard H. Fecher}
\author{Claudia Felser}
\affiliation{Institut f\"{u}r Anorganische Chemie und Analytische Chemie,
             Johannes Gutenberg - Universtit\"{a}t,  55099 Mainz,
             Germany}
\author{Lijun Zhang}
\author{David J. Singh}
\affiliation{Materials Science and Technology Division, Oak Ridge
National Laboratory, Oak Ridge, TN 37831-6114, USA}

\date{\today}

\begin{abstract}

We  report density  functional studies  of the  Fe$_{1-x}$Cu$_x$Se alloy
done  using  supercell  and  coherent potential  approximation  methods.
Magnetic  behavior was  investigated using  the disordered  local moment
approach.  We  find that Cu  occurs in a nominal  $d^{10}$ configuration
and is highly  disruptive to the electronic structure  of the Fe sheets.
This  would be  consistent  with  a metal  insulator  transition due  to
Anderson localization.  We further find  a strong cross over from a weak
moment itinerant  system to a local  moment magnet at  $x \approx 0.12$.
We  associate  this with  the  experimentally  observed  jump near  this
concentration.  Our results are  consistent with the characterization of
this concentration dependent jump as a transition to a spin-glass.

\end{abstract}



\maketitle

\section{introduction}

The  discovery  of high-temperature  superconductivity  in Fe  compounds
\cite{KWHH08} has  led to widespread interest both  in understanding the
superconductivity and in unraveling the normal state properties of these
unusual materials.   These Fe-based materials and cuprates  are the only
known  superconductors  with   critical  temperatures  exceeding  50  K.
Therefore it is natural that  the relationship between these two classes
of materials has  been the focus of much  investigation.  At present, it
seems that the Fe-based materials  are rather different from cuprates at
least from an electronic point of view -- in particular they do not show
Mott insulating phases, and  generally appear to manifest more metallic,
itinerant electron physics than cuprates.
\cite{sebastian,lu-pe,yang-xas}  It is unclear  whether this  means that
the Fe-superconductors are fundamentally different from the cuprates, or
whether there is a more subtle connection that remains to be found.

In this regard, the FeSe  binary provides a particularly useful material
for investigation both because of its relative simplicity and because of
the   rich    variety   of   properties    including   superconductivity
\cite{HLY+08,MTT+08} found  in it, as well  as in the  related FeS, FeTe
and  alloy systems.   FeSe  represents a  robust  superconductor with  a
remarkable  increase of  $T_{\rm  C}$ under  pressure  reported by  many
groups. \cite{MTT+08,HLY+08,MMT+09,MTO+09,GSL+09,IAN+09}   Furthermore,
this system  is chemically amenable  to a wide variety  of substitutions
while   still  admitting   the  growth   of  high   quality  crystalline
samples. The  response of metallic materials to  disorder and scattering
of   various   types   (magnetic,   non-magnetic,  etc.)    induced   by
substitutions is  a potentially very  useful probe of the  robustness of
the metallic  phase and of  the relationship of physical  properties and
itinerant electron physics.

The fact that the  Fe-based superconductors are less strongly correlated
than  cuprates  more  readily  allows  the use  of  standard  electronic
structure methodology to make  connections with experiment.  However, we
note  that there  remain significant  errors in  the  density functional
description  of these  materials, {\em  e.g.} in  the  interplay between
structure  and  magnetism,  likely  due  to  strong  spin  fluctuations.
\cite{mazin-mag,singh-physica}  In  any   case,  similar  to  the  other
Fe-based   superconductors   \cite{singh-du,mazin,BDG08,YLH+08},   first
principles investigations  of Fe chalcogenides  \cite{SZSD08} imply very
similar physics for the pnictide and chalcogenide superconductors.

Stoichiometry  is an important  issue for  FeSe. The  material typically
forms with some  amount of excess Fe, which  occupies a crystallographic
site  outside the  Fe plane.  The formula  may therefore  be  written as
Fe$_{1+x}$Se  or  equivalently  FeSe$_{1-y}$.  Here we  use  the  former
notation to emphasize the additional  Fe site.  There have been attempts
to modulate  the excess Fe by  alloying with Co (by  substitution) or Na
(by intercalation). \cite{LFHJ08}  According to that work, the number of
excess  valence electrons  (i.e. the  experimental doping  level  of the
samples) was ${x\approx 0.027}$ for  Co and ${x\approx 0.03}$ for Na. In
both cases the  $T_{\rm C}$ was approximately 8.3  K.  Other experiments
\cite{IAN+09}  showed that $x=0.01$  is already  sufficient to  keep the
superconductivity. At the  same time the upper value of  $x$ is bound by
oxygen  contamination. \cite{MHK+09}  In another  work  \cite{WMC09} the
same  authors  reported that  the  strongest  superconducting signal  is
observed closer  to stoichiometry ${(x=0.01)}$ and that  at $x=0.01$ the
system  is  much  closer  to  the ideal  tetragonal  phase,  while  less
stoichiometric  samples  have a  weaker  superconducting transition  and
magnetic  contamination.   The fact  that  with  ${x\rightarrow 0}$  the
system  has stronger  superconductivity  and at  the  same time  becomes
magnetically and structurally unstable are indications that there may be
a quantum  critical point (QCP) nearby.   This view is  supported by the
fact that  near ideal stoichiometry Fe$_{1.01}$Se shows  a tetragonal to
orthorhombic distortion  as it is cooled through  90K. \cite{mcqueen} In
other  Fe-based superconductors  this  type of  distortion  occurs as  a
precursor  to  the  spin   density  wave  (SDW)  magnetic  ordering  (or
coincident with the  SDW), but Fe$_{1.01}$Se does not  show SDW ordering
down to the lowest temperature.

In fact,  many authors have discussed the  association between magnetism
and  superconductivity   in  these  materials,   with  superconductivity
appearing when magnetism is suppressed.  \cite{CHL+08,chen-sdw,ma} While
there is  much work pointing at  the possibility of a  QCP affecting the
physics of the Fe-based superconductors, there is not yet an established
consensus regarding its nature.  On the one hand it has been argued that
there is a  nearness to localization driven by  Coulomb interactions and
that this underlies a quantum critical point. \cite{DSZA09} However, the
phase diagrams  do not  show Mott insulators,  and in fact  the magnetic
phases, including  the spin-density  wave (SDW) phase  are unambiguously
metallic,   \cite{sebastian}  suggesting   a   quantum  critical   point
associated with itinerant magnetism.  Experiments probing the details of
the   interplay  between  magnetic   order  and   superconductivity  are
complicated by  the fact that  chemical disorder in doped  samples makes
investigation  of the  critical  point difficult.  As  we discuss  here,
results for FeSe indicate that it  is very close to the magnetic quantum
critical point and therefore may be a very useful system for elucidating
the associated physics.

Before  proceeding  we  mention  previous first  principles  studies  of
defects and doping in iron  chalcogenides.  Regarding excess Fe, Lee and
co-workers early on showed that chalcogen vacancies if present will lead
to  the   formation  of  ferrimagnetic   clusters  in  FeSe   and  FeTe,
\cite{LPP08}  while Zhang  and co-workers  found that  excess Fe  in the
Fe$_{1+x}$Te  system  serves both  as  an  electron  dopant and  also  a
magnetic  impurity.  Han  and Savrasov  \cite{HS09} reported  a detailed
analysis of the magnetic instabilities and magnetic ordering in relation
to the Fermi surface as a function of excess Fe concentration.

The  present  work is  motivated  by  recent  experimental findings  for
Cu-substituted Fe$_{1.01}$Se. \cite{WMK+09}  In particular, it was found
that Cu substitutes  for Fe in the  Fe plane, and that there  is a rapid
suppression of  $T_{\rm C}$ in  the concentration range $0-4\,\%$  and a
subsequent metal--insulator  transition at $  \sim 4\,\%$.  Furthermore,
there is  a development of  dynamical magnetic fluctuations  detected by
NMR,  which noticeably  rise at  $\sim$ 12\,\%  of Cu  substitution.  We
begin  with supercell  calculations addressing  the chemistry  and local
electronic  structure, and  then proceed  to investigate  the electronic
structure in  more detail  using coherent potential  approximation (CPA)
calculations for the disordered alloy.

\section{Supercell Calculations}

\begin{figure}[htbd]
  \begin{minipage}[c]{1.0\linewidth}
    \includegraphics[width=0.75\textwidth,clip]{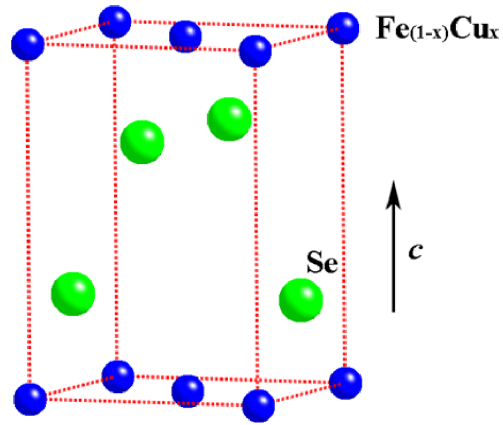}
  \end{minipage}
  \caption{(color online) Crystal structure of Fe$_{1-x}$Cu$_x$Se.}
  \label{FIG:STR}
\end{figure}

First   principles   calculations  were   performed   using  2x2x1   and
2$\sqrt{2}$x2$\sqrt{2}$x1  supercells  based  on  tetragonal  FeSe  (see
Fig.~\ref{FIG:STR}) with one Fe replaced  by Cu.  This corresponds to Cu
concentrations, $x$  in Fe$_{1-x}$Cu$_x$Se of  $x$=0.125 and $x$=0.0625,
respectively.   These  calculations  were  done  using  the  generalized
gradient approximation of Perdew, Burke and Ernzerhof (PBE). \cite{pbe}
The lattice parameters of the  supercells were fixed to the experimental
values     for    $x$=0.12     and    $x$=0.06     as     reported    in
Ref.~\onlinecite{WMK+09}. However, the  atomic positions within the cell
were determined  by energy minimization. This  structural relaxation was
performed using the projector augmented wave (PAW) method as implemented
in the  VASP code, \cite{vasp}  with an energy  cut-off of 350  eV.  The
relaxed structures  showed some expansion  of the lattice mainly  in the
in-plane direction around the Cu impurity sites.  The relaxed Cu-Se bond
lengths  were $\sim$  2.44 \AA,  as compared  to Fe-Se  bond  lengths of
2.25\AA-2.33\AA.   This  is consistent  with  the  increase in  in-plane
lattice parameter  seen experimentally. \cite{WMK+09,li}  The electronic
structures  were calculated  using  the more  precise general  potential
linearized augmented plane-wave (LAPW) method, \cite{singh_book} with the
augmented plane wave plus local orbital (APW+LO) implementation
\cite{sjostedt} of the WIEN2k code. \cite{wien} LAPW sphere radii of 
2.1\,$a_0$ for all sites, with well converged basis sets and zone samplings
were  used in this  calculation.  The  consistency of  the PAW  and LAPW
calculations was checked via the LAPW forces for the structures obtained
via total  energy relaxation with the  PAW method. The  maximum force in
the LAPW calculation was 2\,mRy/$a_0$.

\begin{figure}[htbd]
  \begin{minipage}[c]{1.0\linewidth}
    \includegraphics[width=0.75\textwidth,clip]{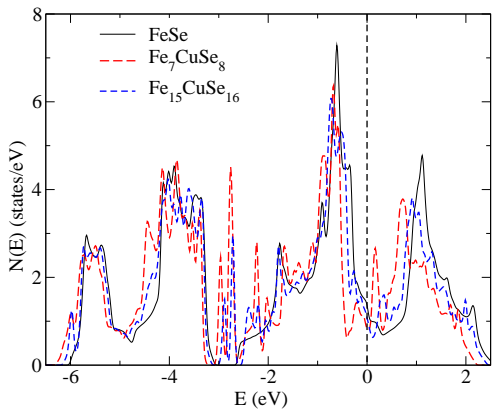}
  \end{minipage}
  \caption{(color online)
Total electronic density of states for Fe$_{1-x}$Cu$_x$Se
supercells on a per formula unit basis. The Fermi energy is at 0 eV.}
  \label{FIG:dos-total}
\end{figure}

\begin{figure}[htbd]
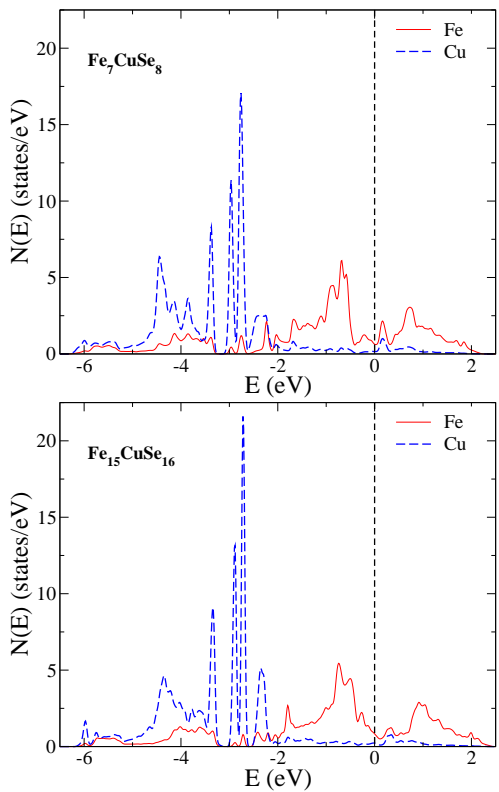

  \begin{minipage}[c]{1.0\linewidth}
    \includegraphics[width=0.75\textwidth,clip]{figure_3a.eps}
    \includegraphics[width=0.75\textwidth,clip]{figure_3b.eps}
  \end{minipage}
  \caption{(color online)
Fe and Cu $d$ contributions to the density of states for the $x$=0.125
(top) and $x$=0.0625 (bottom) supercells. These are given by projections
onto the LAPW spheres for Cu and averaged over the Fe sites, on a per atom
basis.}
  \label{FIG:dos-cufe}
\end{figure}

The  electronic density of  states for  the two  supercells is  shown in
Fig.~\ref{FIG:dos-total}, while the average  Fe and Cu $d$ contributions
as  defined  by   projections  onto  the  LAPW  spheres   are  shown  in
Fig.~\ref{FIG:dos-cufe}.  As may be seen  the Cu $d$ bands are at ${\sim
3}$~eV binding  energy, and are  therefore fully occupied for  a nominal
$d^{10}$ Cu configuration.  We also note that there are shifts in the Fe
density of states as seen in Fig.~\ref{FIG:dos-total}.  These shifts are
consistent  with electron  doping.  However,  substitution of  a nominal
Fe$^{2+}$ by Cu$^{1+}$ would ordinarily  be expected to lead to hole and
not electron doping.  This suggests a bonding rearrangement involving Se
around the Cu site.  There is also some Cu $d$ contribution to the bands
above  the  Fermi level  $E_F$  indicating  covalency  involving Cu  $d$
states.   However,  the  magnitude (see  Fig.~\ref{FIG:dos-cufe})  seems
inadequate to explain the  observed electron doping, implying that Se-Se
bonding  may  be important  around  the Cu  site.   This  is similar  to
findings  for  CuSe  in  various  structural  modifications  showing  an
interplay of different bonding types. \cite{milman}

In any  case, from the  positions of the  Cu $d$ states at  high binding
energy  it is clear  that Cu  substitution is  highly disruptive  to the
electronic structure of  the Fe sheets in FeSe.  This  is in contrast to
the  case of Co  in the  arsenides, which  leads to  the formation  of a
coherent electronic structure with effective doping. \cite{sefat}  As
such,  Cu  alloying  cannot  be  considered  with  the  virtual  crystal
approximation, and  more sophisticated  treatments such as  the coherent
potential   approximation  (CPA)  \cite{Gyo72,SW84,But85}   are  needed.
Considering the  strong scattering implied by  this result, localization
could  be caused  by  disorder, i.e.  Anderson localization,  \cite{lee}
should  be  considered in  the  context  of  the insulating  phase  that
develops  at Cu  concentrations  above  4\%.  Finally,  in  view of  the
proximity  of these phases  to magnetism,  disruption of  the electronic
structure  by  scattering  (which  works against  itinerancy)  would  be
expected  to  lead to  the  formation of  local  moments  around the  Cu
sites. These would  be distributed over the Fe atoms  around the Cu, but
because of  the $d^{10}$ configuration would  not exist on  the Cu atoms
themselves. \cite{moment-note}

\section{Coherent Potential Approximation Calculations}

Our  CPA  calculations were  performed  using  the tetragonal  structure
(Space  group:~115) (see  Fig.~\ref{FIG:STR}),  with lattice  parameters
from  experiment, \cite{WMK+09}  similar to  the  supercell calculations
described   above.   Additionally,   in  these   calculations   we  used
experimental atomic positions.  It follows that with Cu substitution the
system linearly  expands in the  plane containing Fe atoms  and linearly
shrinks in the perpendicular direction. The volume increases weakly with
Cu content.  At  the same time, the position of the  Se atoms (not fixed
by the point  group symmetry), remains practically constant  and is here
set    to    $z=0.2707$   (in    multiples    of    the   $c$    lattice
parameter). Interestingly, by extrapolating the lattice parameters up to
50~\%  Cu  one  arrives very  close  to  the  structure  of one  of  the
modifications of CuFeSe$_2$. \cite{DMP94} That material is a
semiconductor with low-temperature magnetism. \cite{DMP94,woolley}

\begin{figure}[htbd]
  \begin{minipage}[c]{1.0\linewidth}
    \includegraphics[width=0.75\textwidth,clip]{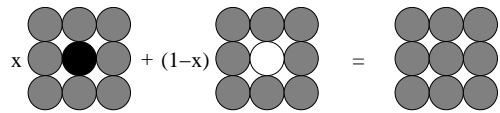}
  \end{minipage}
  \caption{The CPA  model: the Green's function of  the effective medium
(gray atoms) is obtained as an average of the partial impurities Green's
functions (black and white atoms).}
  \label{FIG:CPA}
\end{figure}

The  idea of the  CPA is  to replace  the random  array of  real on-site
potentials by  an ordered array of effective  potentials. The scattering
properties   of   the    effective   potential   are   than   determined
self-consistently  in terms  of  the local  mean-field  theory with  the
condition that the  total Green's function of the  effective system does
not change upon  replacement of the single effective  potential with the
real one. This idea is sketched in Fig.~\ref{FIG:CPA}.

The computational  cost of CPA calculations is  significantly lower than
supercell approaches  where the disorder  is approximated by  a randomly
generated configurations within a  large number of large supercells. The
important advantage  is that the  CPA does not affect  the translational
(as  well as  the point  group) symmetry  of the  unit cell  whereas the
supercells must be  chosen sufficiently large to avoid  effects from the
assumed  order or  periodic images,  which are  caused by  an artificial
translational symmetry.  The CPA allows investigation  of the electronic
structure as a  continuous function of the substitution  level, which is
very important describing  phase transitions as well as  in studying the
evolution of  the electron structure  with concentration.  The  CPA also
provides   type-resolved   contributions    of   the   different   local
quantities.  The disadvantage  is  that as  any  mean-field theory,  the
standard  CPA does  not include  the local  environment effects  such as
preferential ordering, \cite{ME06} the Invar effect \cite{BOG+97,CEE+02}
and   lattice   relaxations  around   the   impurity   site.   The   low
concentrations  of  interest here,  and  the  results  of the  supercell
calculations,  which find  modest changes  in local  structure, indicate
that the effects of local ordering and lattice distortion are not likely
to be the crucial factors in determining the electronic structure of the
alloy.

\begin{figure}[htbd]
  \begin{minipage}[c]{1.0\linewidth}
    \includegraphics[width=0.75\textwidth,clip]{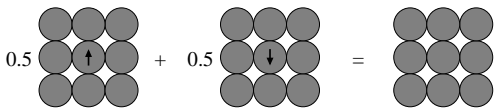}
  \end{minipage}
  \caption{The  DLM  model:  the   Green's  function  of  the  effective
non-magnetic medium is  obtained as an average of  the partial polarized
impurity Green's functions.}
  \label{FIG:DLM}
\end{figure}

The Fe$_{1-x}$Cu$_x$Se system is known  to be non-magnetic, in the sense
of  not having  ordered magnetism,  over  the investigated  range of  Cu
concentrations.  However,  as mentioned, there  is experimental evidence
that magnetic  fluctuations noticeably strengthen  above 12\,\%~Cu.  The
consistent  way  of studying  these  fluctuations  is  to determine  the
coefficients  $A_i$ of  the Ginzburg-Landau  energy  expansion $E=E_{\rm
PM}+\sum_{i}A_{i}M^{2i}$  in the  local magnetization  $M$.  \cite{RJ97}
The  best  local  approximation  to the  disordered  paramagnetic  state
$E_{\rm PM}$  is given by  the so-called {\it disordered  local moments}
(DLM)  approach,  \cite{PSSW83,GPS+85}  which  is used  in  the  present
work. The  DLM effective  medium gives an  accurate representation  of a
paramagnetic state  with randomly oriented spins by  using an equiatomic
random alloy of spin-up and  spin-down atoms, in complete analogy to the
CPA (see Fig.~\ref{FIG:DLM}).

In  this  context the  DLM  picture provides  a  common  basis to  study
different types of possible magnetic order. The important feature of the
DLM  is the  consistent description  of the  paramagnetic state  for the
systems  with  itinerant  and local  moment  behavior;\cite{RJ97,RKMJ07}
local moment systems keep their finite moments in the paramagnetic state
whereas  itinerant systems become  locally non-magnetic.   The essential
assumption of the DLM is  the quasi-static treatment of fluctuations. It
assumes  that the  fluctuations of  the local  magnetic moment  are much
slower  than  the  electron  motion,  so  that  the  instantaneous  band
structure corresponds  to the equilibrium electronic  state. This should
lead to a  certain overestimation in case of  large fluctuating magnetic
moments  and to  a  certain suppression  in  case of  small moments,  by
sharply emphasizing magnetic transitions.

Here  we  use the  KKR  (Korringa-Kohn-Rostoker)  method  in the  Munich
SPR-KKR  (spin-polarized relativistic) package  \cite{EB96} for  our CPA
calculations.  These calculations were  done with the local spin-density
approximation (LSDA) in the form of Vosko, Wilk and Nusair \cite{VWN80}.
Here we use  the full-symmetry potential (FP) method  to account for the
non-spherical  contributions  to the  one-electron  potential.  This  is
important because the layered structure of FeSe is strongly anisotropic.
As expected, relativistic effects are found to be relatively small. Thus
although the  calculations are fully  relativistic, only spin  moment is
reported  in  the following  as  the orbital  moments  are  found to  be
negligibly small.

The   effective    DLM   medium    is   represented   by    the   random
Fe$_{0.5}$Fe$_{0.5}$  alloy, with  ${\pm~2\,\mu_{\rm  B}}$ anti-parallel
spins initially induced  for each on-site component. The  system is then
allowed to  relax during the self-consistent iteration  process. In case
of locally  non-magnetic solutions  the proximity of  the system  to the
magnetic  state   is  estimated  by   considering  the  Fe   local  spin
susceptibility   ${\chi=\left.d\mu_{\rm}/dB_{\rm  ext}\right|_{\mu=0}}$.
This is calculated  by applying the small external  local magnetic field
of $dB_{\rm ext}=2$~mRy.

The  symmetry  of  the   magnetic  fluctuations  are  considered  to  be
characterized  by  the  exchange  coupling  constants  $J_{ij}$  of  the
Heisenberg         local        moments         picture:        ${H_{\rm
eff}=-\sum_{i>j}J_{ij}\vec{\mu}_i\vec{\mu}_j}$, where $\vec{\mu}_{i}$ is
the local magnetic moment of the $i$-th atomic site.  However, it should
be emphasized that the use of a Heisenberg model in analyzing low energy
magnetic fluctuations does not necessarily imply local moment magnetism.
Here,    the     so-called    real-space    formalism     is    utilized
\cite{LKAG87}. Again,  in case of the non-magnetic  solution, a magnetic
moment is induced by a small external field of $2$~mRy.

\section{CPA Results}

\subsection*{Magnetic moments and susceptibility}

\begin{figure}[htbd]
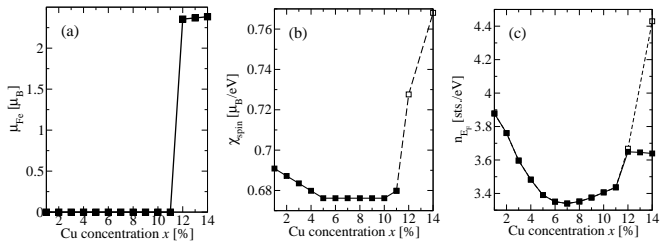

  \begin{minipage}[c]{0.32\linewidth}
    \includegraphics[width=1.0\textwidth,clip]{figure_6a.eps}
  \end{minipage}~
  \begin{minipage}[c]{0.33\linewidth}
    \includegraphics[width=1.0\textwidth,clip]{figure_6b.eps}
  \end{minipage}~
  \begin{minipage}[c]{0.32\linewidth}
    \includegraphics[width=1.0\textwidth,clip]{figure_6c.eps}
  \end{minipage}
  \caption{Local magnetic  moments of Fe (a),  local spin susceptibility
(b) and the  total DOS at the Fermi energy (c)  calculated as a function
of  Cu  concentration.   Filled  squares  mark  DLM,  hollow  squares  -
non-magnetic calculations.}
  \label{FIG:MOM-CHI-DOS}
\end{figure}

Regardless of the calculation  mode (magnetic or non-magnetic, fully- or
non-relativistic,  spherical or non-spherical  potential), we  find that
all properties  (density of states  (DOS) at the Fermi  energy, magnetic
moments, susceptibility,  etc.) calculated as a function  of $x$ exhibit
discontinuities    or   jumps   at    a   Cu    concentration,   $x_{\rm
m}\,\approx\,12\%$.    Importantly,   this   concentration   corresponds
remarkably well  to the experimentally  observed onset of  the dynamical
magnetic fluctuations.  \cite{WMK+09}  The DLM calculations  lead to the
appearance of  local moments at  the Fe site  that vary weakly  within a
given   regime:   low  Cu-concentration   (${x<x_{\rm   m}}$)  or   high
Cu-concentration (${x>x_{\rm m}}$) (Fig.~\ref{FIG:MOM-CHI-DOS}(a)).

Within the low Cu-concentration regime the magnetic moment is suppressed
almost to zero (${\sim  10^{-3}\,\mu_{\rm B}}$), however at ${x=x_m}$ it
rises  sharply  to  about  ${2.35~\mu_{\rm  B}}$.   The  Fe  local  spin
susceptibility  shows  a similar  discontinuous  behavior at  ${x=x_{\rm
m}}$,  as shown  in Fig.~\ref{FIG:MOM-CHI-DOS}(b).   On the  other hand,
besides  the sharp  magnetic  transition at  $x_{\rm  m}$, the  magnetic
susceptibility indicates that the system approaches magnetism in the low
Cu-concentration  regime  as  well,  however  its  ground-state  remains
non-magnetic.   As  may  be  seen  (Fig.~\ref{FIG:MOM-CHI-DOS}(c)),  the
changes in  magnetic behavior  as a function  of $x$ closely  follow the
behavior of the  density of states at the Fermi  energy and is therefore
related to the band structure.

We emphasize that the  DLM calculations provide information about moment
formation  in the  disordered (paramagnetic)  state, not  directly about
magnetic ordering.  In fact, at the  LSDA or GGA levels  FeSe is already
unstable   against   magnetic   ordering,   specifically   against   SDW
order.   \cite{SZSD08}  This  ordering   is  presumably   suppressed  by
spin-fluctuations the  detailed nature of which is  not fully understood
at   present,  but  which   might  be   central  in   understanding  the
superconductivity   of  these   phases.   What   a  cross-over   from  a
non-magnetic  to  a  state  with  moments  in the  DLM  indicates  is  a
cross-over from  a state  that may have  itinerant magnetism to  a state
with stable moments independent of ordering.  This is therefore a change
in  the nature of  the magnetism  at this  value of  $x$.  While  such a
cross-over  is not  directly a  QCP it  may be  an indicator.   FeSe, as
mentioned,  does  not  show  long  range magnetic  order,  but  magnetic
ordering is very difficult to  avoid in clean metals with stable moments
and   electronic  structures  that   are  neither   low-dimensional  nor
geometrically  frustrated.  The  effect   of  the  disorder  induced  by
relatively low amounts of Cu alloying appears to be a change of magnetic
character from a softer moment  (itinerant) regime to a regime with more
robust local moments.
 
\subsection*{Exchange coupling}

\begin{figure}[htbd]
  \begin{minipage}[c]{1.0\linewidth}
    \includegraphics[width=1.0\textwidth,clip]{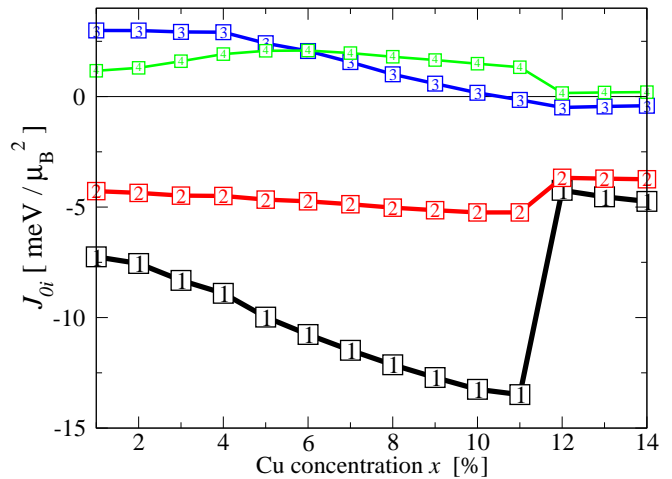}
  \end{minipage}
  \caption{(Color online) Exchange  constants $J_{oi}$ ($i=1,2,3,4$) for
the  4   nearest-neighbor  in-plane  couplings  as  a   function  of  Cu
concentration. Each  $J_{0i}$ coupling curve is  marked by corresponding
index $i$.}
  \label{FIG:JXC}
\end{figure}

Turning  to  the exchange  couplings,  we  find  very small  inter-plane
interactions below the precision  of the calculations.  Furthermore, the
in-plane contributions are significant  only for the first four in-plane
coordination circles.  The corresponding values are shown  as a function
of Cu concentration in Fig.~\ref{FIG:JXC}.

Since in the range of  $x<x_{\rm m}$, $J_{01}$ has larger amplitude than
$J_{02}$, while the number of the first- and second-nearest neighbors is
the  same,  the  nearest   antiferromagnetic  order  seems  to  be  more
preferable.  For the pure FeSe  the particular order was already studied
in more  details by other  authors. \cite{SZSD08,LCG+09,MJH+09,HS09}  In
the  extrapolation  from  the  low Cu-concentration  limit  the  present
exchange  coupling  constants agree  well  with  those published  before.
\cite{SZSD08,LCG+09} On  the other  hand, as we  have seen,  the system
does not have stable magnetic moments within $x<x_{\rm m}$.

The monotonously  increasing amplitude of  the nearest-neighbor coupling
indicates  the  growing   electron  delocalization  with  increasing  Cu
content. However, at ${x=x_{\rm m}}$ one observes a rapid suppression of
all $J_{ij}$,  which in  turn indicates the  sudden localization  of the
$d$-electrons and  corresponds to the  development of the  strong atomic
moments. In this regime ${J_{01}\approx J_{02}}$, while the higher-order
coupling  constants nearly  vanish.   This situation  with stable  local
moments emerging  at $x \approx 0.12$,  competing exchange interactions,
and disordered Cu  positions (which correspond to sites  with no moment)
is consistent with the formation of a spin-glass.

\section{Band structure analysis}

\subsection*{Band structure}

\begin{figure}[htbd]
  \begin{minipage}[c]{0.9\linewidth}
    \includegraphics[width=0.75\textwidth,clip]{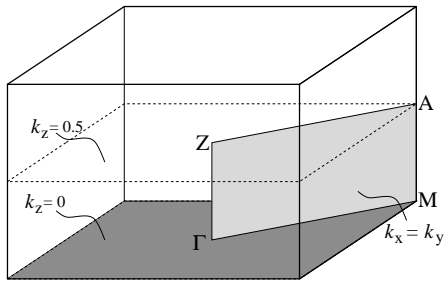}
  \end{minipage}
  \caption{The BZ scheme. The reciprocal $(k_x,k_y,k_z)$-coordinates are
given  in units  of  $(2\pi/a,2\pi/a,2\pi/c)$.  The  marked details  are
discussed in the text.}
  \label{FIG:BZ}
\end{figure}

\begin{figure}[htbd]
  \begin{minipage}[c]{0.9\linewidth}
    \includegraphics[width=1.0\textwidth,clip]{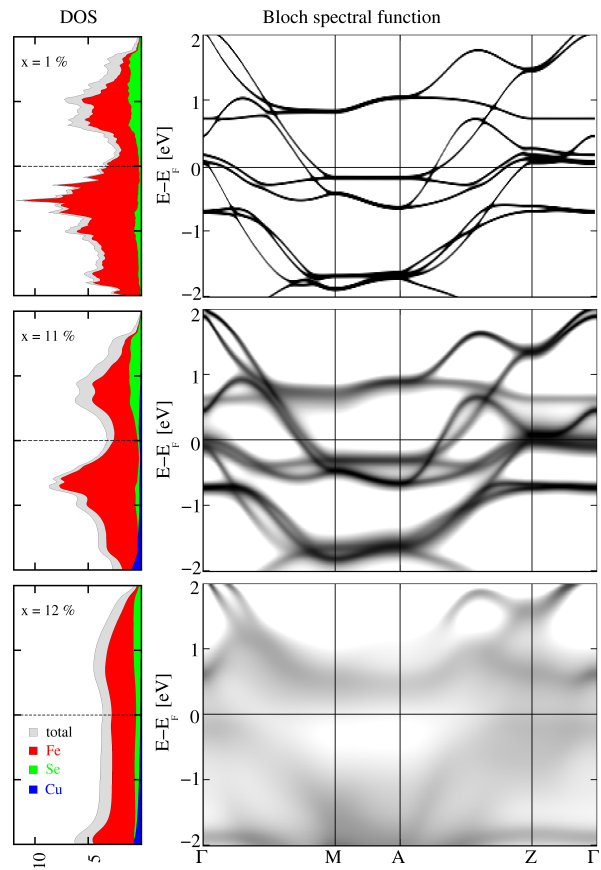}
  \end{minipage}
  \caption{(Color online) Atomic type-resolved DOS and
the corresponding Bloch spectral functions
       for ${x=1, 11, 12\,\%}$ as calculated within the DLM.}
  \label{FIG:BNS-LDA-FD-DM}
\end{figure}

In the  following we consider  the Bloch spectral functions,  which take
the  role of the  band structure  in the  disordered system.   These are
plotted along  the path shown in Fig.~\ref{FIG:BZ}.   The calculated DOS
and    the    corresponding   spectral    functions    are   shown    in
Fig.~\ref{FIG:BNS-LDA-FD-DM}.    The   general    trend   in   the   low
Cu-concentration  regime is  the redistribution  of the  spectral weight
from the  M-A onto  the Z-$\Gamma$ line,  by shifting the  whole picture
down in energy with respect  to the Fermi level.  Qualitatively, this is
exactly what was  found in our supercell calculations  where in spite of
the Cu $d^{10}$ configuration, Cu substitution produces electron doping.

As in  the other Fe-based  superconductors the nearness to  magnetism of
FeSe  is  related to  the  electronic  structure  specifically the  high
density  of states derived  from Fe  $d$ orbitals  and the  nested Fermi
surface.   The Fermi  energy at  $x$=0 occurs  towards the  bottom  of a
pseudo-gap in the DOS. As seen in the Bloch spectral functions there are
two effects as  Cu is alloyed into the material. First  of all the Fermi
energy is raised and secondly, as a consequence of the disorder, the DOS
as  a function of  energy is  smoothed, filling  in the  pseudogap. This
explains the initial  decrease, followed by an increase  in the tendency
towards moment formation.  When the Cu concentration reaches $x_{\rm m}$
value there  is a dramatic change  in the spectral  function.  Since the
system develops  large local moments, the  DLM spin disorder  leads to a
strong incoherence in the bands.

\subsection*{Fermi surface}

To  study   the  nesting  effects  we   track  the  substitution-induced
transformation of the Fermi surface in  the BZ. In the following we will
consider  the  Fermi  surface  cross-sections by  horizontal  (${k_z=\rm
const}$) and vertical (${k_x=k_y}$) planes.

\begin{figure}[htdb] 
  \begin{minipage}[c]{0.9\linewidth}
    \includegraphics[width=1.0\textwidth,clip]{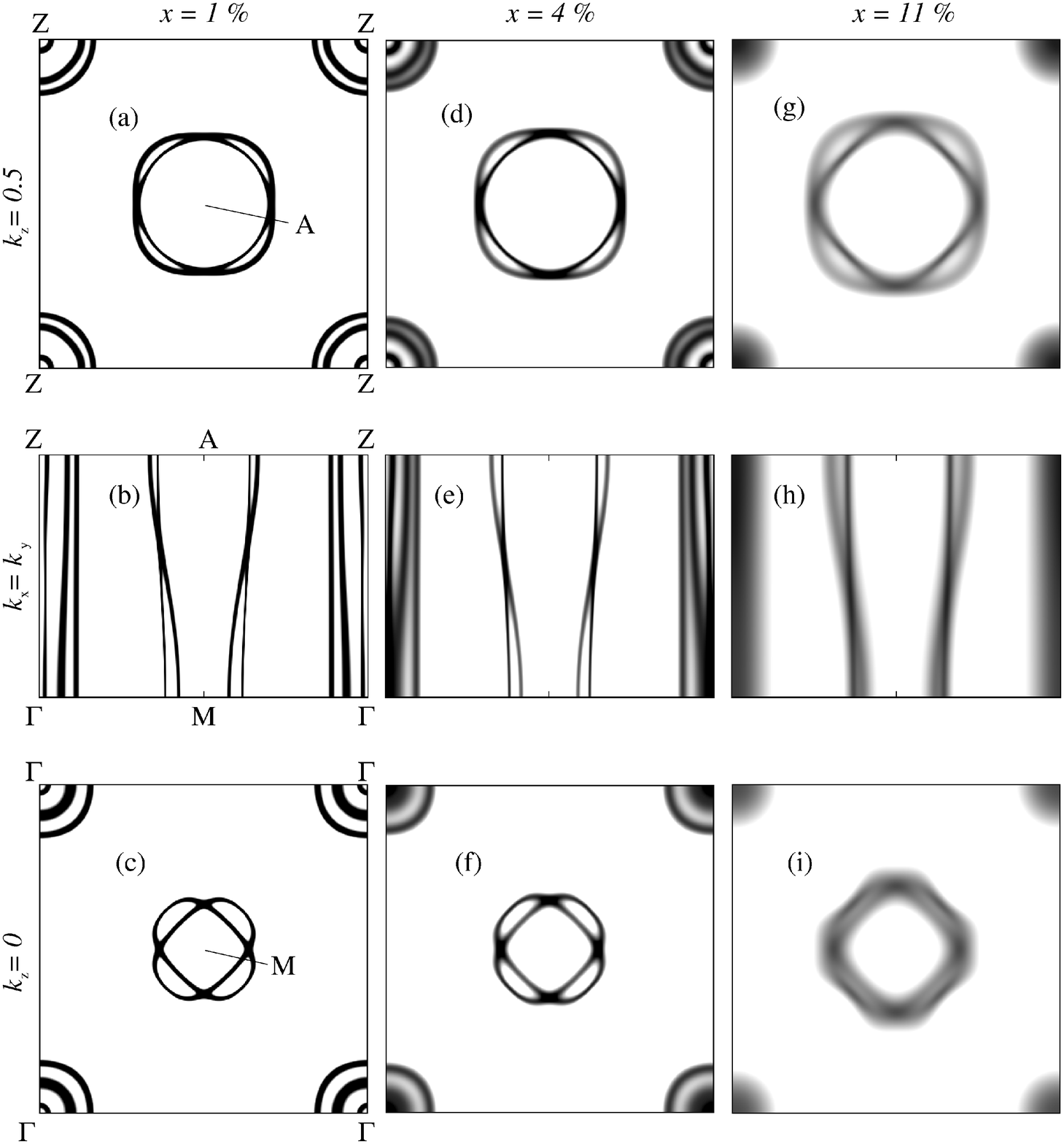}
  \end{minipage}
  \caption{Cross-sections of the Fermi surface in BZ by  ${k_z=0.5}$ (a,
    d, g), ${k_x=k_y}$ (b, e, h) and ${k_z=0}$ (c, f, i) planes.}
  \label{FIG:FSC-LDA-FD-DM}
\end{figure}

In general,  the electronic  structures of the  Fe-based superconductors
show disconnected Fermi surfaces  consisting of hole sections around the
zone center  (along $\Gamma$-Z)  and two electron  sections at  the zone
corner   (along   M-A   direction).   Our   calculated   Fermi   surface
cross-sections (Fig.~\ref{FIG:FSC-LDA-FD-DM}) are in good agreement with
this picture.

It  turns  out  that  the  growing  chemical  disorder  with  increasing
Cu-concentration  destroys the Fermi  surface, especially  affecting the
hole-pockets (centered at $\Gamma$-Z). Since the in-plane hole--electron
nesting  is the  main mechanism  for the  superconductivity in  FeSe and
related  compounds,   \cite{SZSD08}  the  superconducting   signal  must
strongly  attenuate.  Indeed, the  experimentally  measured $T_{\rm  C}$
falls  down  very  fast  and  completely vanishes  at  ${x\approx 4}$~\%.
\cite{WMK+09}

\begin{figure}[htbd]
  \begin{minipage}[c]{0.9\linewidth}
    \includegraphics[width=1.0\textwidth,clip]{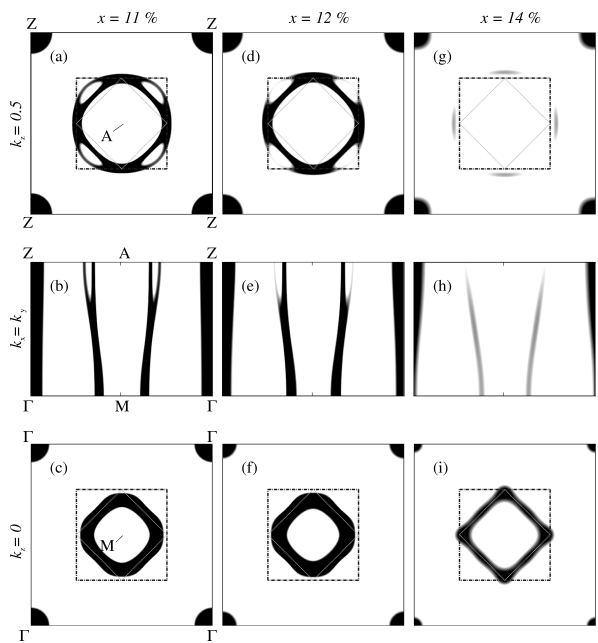}
  \end{minipage}
  \caption{Cross-sections of the Fermi  surface in BZ by ${k_z=0.5}$ (a,
d,   g),   ${k_x=k_y}$   (b,   e,   h)   and   ${k_z=0}$   (c,   f,   i)
planes. Calculations are done within the non-magnetic regime.}
  \label{FIG:FSC-LDA-FD-DM-SHARP}
\end{figure}

Although the  reason for  the magnetic transition  can be  understood in
terms of the band  structure (Fig.~\ref{FIG:BNS-LDA-FD-DM}), it is still
useful  to observe the  accompanying changes  of the  non-magnetic Fermi
surface.  The  corresponding results for 11\,\%  and forced non-magnetic
calculations     for     12    and     14\,\%     are    compared     on
Fig.~\ref{FIG:FSC-LDA-FD-DM-SHARP}.  Here  the low-weight  contributions
are  cut  off,  otherwise  the  band structure  analysis  will  be  much
complicated by chemical disorder.

Thus,  the Fermi  weight rapidly  ``spills'' from  ${k_z=0.5}$  onto the
${k_z=0}$  plane.   This  shows a  loss  of  the  Fermi surface  in  the
${k_z=0.5}$ plane first as a function of Cu alloying although there is a
strong loss of  spectral weight at all $k_z$  consistent with insulating
character.  At  the same time  the central figure  does not blow  up any
more, but  rather pulls the  borders of the  magnetic BZ (MBZ)  (the two
relevant in-plane MBZs are marked by  the dashed and the dotted lines on
Fig.~\ref{FIG:FSC-LDA-FD-DM-SHARP}).   Analogous behavior was  shown for
the  related Fe$_{1+x}$Te  compound\cite{HS09}:  it was  suggested that
such  redistribution of the  Fermi weight  between two  neighboring MBZs
reflects  the  competition between  magnetic  interactions of  different
symmetries. The inhomogeneity  becomes especially sharp at ${x=14\,\%}$,
indicating  that  the  system  starts  to prefer  the  certain  in-plane
magnetic order for ${x>x_{\rm m}}$.

Preferential  accumulation of  the  Fermi  weight on  the  edges of  the
corresponding MBZ in the ${k_z=0}$ plane indicates the delocalization of
the  conduction electrons  in real  space.  This  delocalization becomes
unfavorable,   since  the  in-plane   lattice  spacing   increases  (the
accompanying vertical real-space delocalization is a minor effect). At a
certain point  ${(x=x_{\rm m})}$  the system undergoes  the localization
transition.  As a result, the  strong magnetic moment arises. This rapid
localization is also reflected by  the reduced amplitude of the exchange
coupling constants for ${x>x_{\rm m}}$ as discussed above.

\section{Summary and conclusions}

To summarize,  we find that Cu  occurs in a  $d^{10}$ configuration when
alloying  into FeSe.  Nonetheless  it serves  as  an effective  electron
dopant.  Importantly,  it is  also a source  of strong  scattering.  The
calculated changes in the Bloch  spectral function with alloying show no
signs  of gapping  near the  Fermi  level over  the concentration  range
studies.  This situation  suggests that  the insulating  phase occurring
above $\sim$  4\% Cu  may be an  Anderson localized system  arising from
disorder  rather  than   a  conventional  semiconductor.   The  magnetic
instability  observed  at  about  $12\,\%$  Cu  is  characterized  as  a
spin-state like transition,  where a soft moment near  magnetic state at
low $x$  gives way to  a state with  stable disordered local  moments at
high  $x$.   This is  consistent  with  the  formation of  a  spin-glass
especially considering the near compensation of nearest and next nearest
exchange   interactions.    This   is   in  accord   with   experimental
observations  \cite{WMK+09}.

The result that the Fermi  weight for the electron sections at $k_z=0.5$
is  suppressed  before  that  at  $k=0$  suggests  a  three  dimensional
character  to the  electronic  structure and  scattering  above $x  \sim
0.1$.  It will  be  quite interesting  to  investigate this  possibility
experimentally,  via  single  crystal  transport  or  experiments  under
strain, especially uniaxial strain if this becomes feasible.

Finally,  we note  that the  present calculations  assume  full disorder
between Cu and Fe,  and also that both Fe and Cu  remain in the Fe plane
and  do not  occupy  other sites,  such as  the  excess Fe  site in  the
interstitial.  It will  be  useful to  test  the extent  to which  these
assumptions hold for the real material, for example by using diffraction
to refine the occupancy of the interstitial site.

\begin{acknowledgments} The  authors are much indebted  to Roser Valenti
and J\"urgen K\"ubler for  helpful discussions. The financial support by
the  SFB/TRR49 is gratefully acknowledged.  Work  at Oak  Ridge National
Laboratory  was  supported by  the  Department  of  Energy, Division  of
Materials Sciences and Engineering.
\end{acknowledgments}


\begin{thebibliography}{55}
\bibitem[{mom()}]{moment-note}
\bibinfo{note}{Another interesting possibility is that moments could form on
  the neighboring Fe but that these could form singlets where the four Fe atoms
  around the Cu could have cancelling moments due to antiferromagnetic
  interactions between them. Recalling that the four Fe atoms around a Cu
  substitutional are second neighbors on the Fe square lattice, the case where
  the Fe moments are aligned would correspond locally to the checkerboard
  (nearest neighbor antiferromagnetic) order, while the antiferromagnetic state
  with cancelling Fe moments would correspond locally to the SDW pattern.}
\end{thebibliography}
\end{document}